\documentclass[twocolumn,a4paper,10pt]{revtex4-1}
\usepackage{CJKutf8,upgreek,fancyhdr}
\usepackage{graphicx}
\usepackage{diagbox}
\usepackage{booktabs}
\usepackage{tabularx}
\usepackage{verbatim}
\usepackage{dcolumn}
\usepackage{multirow}
\usepackage{amssymb,bm,mathrsfs,bbm,amscd}
\usepackage[tbtags]{amsmath}
\usepackage{lastpage}

\begin{document}
\begin{CJK}{UTF8}{gbsn}



\title{The compensation for the edge focusing of chicane bump magnets by harmonic injection in CSNS/RCS}
\author{Xiaohan Lu$^{1,2,3}$}
\author{Shouyan Xu$^{1,3}$}
\author{Sheng Wang$^{1,3}$}

\email{wangs@ihep.ac.cn}
\affiliation{%
$^1$Dongguan Institute of Neutron Science, Dongguan 523808, China}
\affiliation{$^2$University of Chinese Academy of Sciences, Beijing 100049, China}
\affiliation{$^3$Institute of High Energy Physics, Chinese Academy of Sciences, Beijing 100049, China\\}

\begin{abstract}
In the Rapid Cycling Synchrotron(RCS) of China Spallation Neutron Source(CSNS), transverse painting injection is employed to suppress the space-charge effects. The beta-beating caused by edge focusing of the injection bump magnets leads to tune shift and shrinkage of the acceptance, which may result in additional beam loss. In RCS, the main quadrupoles are excited by White resonant power supplies, and the exciting current cannot be arbitrarily programed. Generally, this kind of perturbation could  be corrected by trim-quadrupole, however, we don't have trim-quadrupole in CSNS/RCS. A new method based on the harmonic injection is firstly introduced to compensate the beta-beating caused by edge focusing of the chicane bump magnets at RCS. In this paper, the principle and feasibility of compensation scheme were presented. The simulation study was done on the application of the new method to the CSNS/RCS, and the results show the validity and effectiveness of the method.
\begin{description}
\item[PACS numbers]{29.27.Ac, 29.29.Bd, 41.85.Lc}
\end{description}
\end{abstract}

\maketitle

\section{Introduction}
\begin{figure}[htp]
\includegraphics[width=8cm]{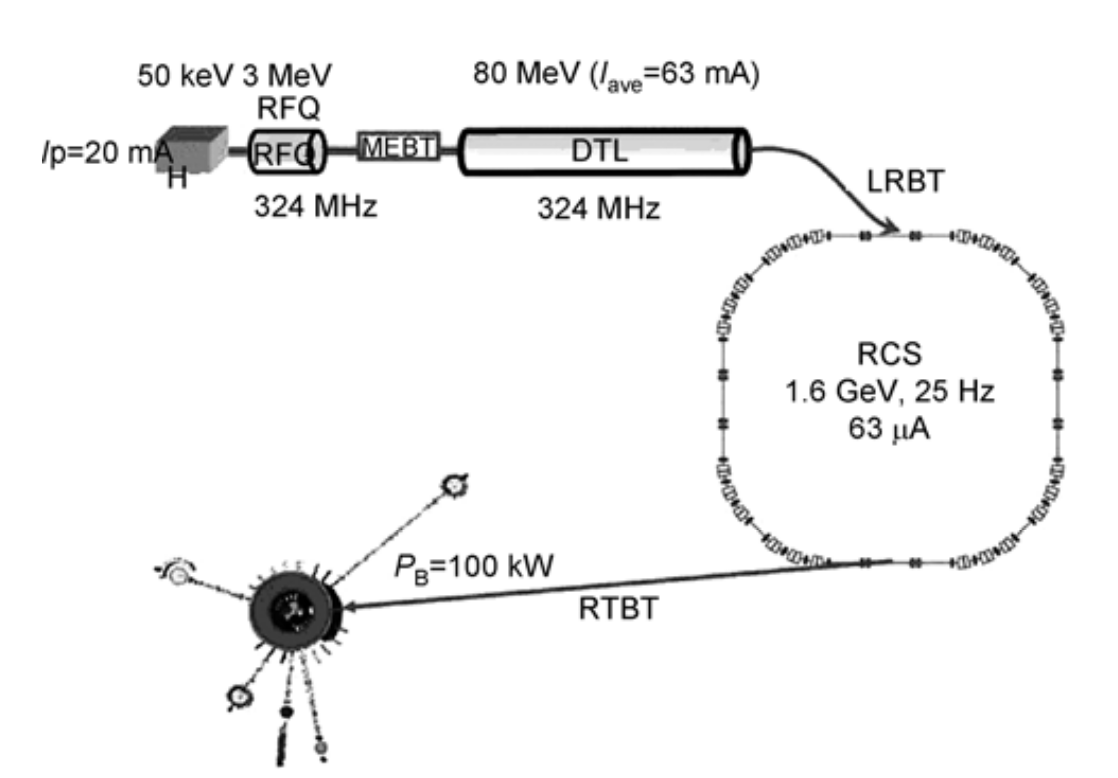}
\caption{\label{fig1} The schematic layout of CSNS}
\end{figure}

China Spallation Neutron Source (CSNS) consists of an 80 MeV linac and a 1.6 GeV Rapid Cycling Synchrotron (RCS) with repetition rate of 25 Hz~\citep{ref1}. As shown in Fig.~\ref{fig1}, an 80-MeV $H^-$ beam is delivered from linac to the RCS injection point and stripped to proton by stripping foil. RCS accelerates the protons up to the designed energy of 1.6 GeV and extracts the beam with the power of 100 kW to the neutron target~\citep{ref2}. 

\begin{figure}[htb]
\includegraphics[width=8cm,height=5.5cm]{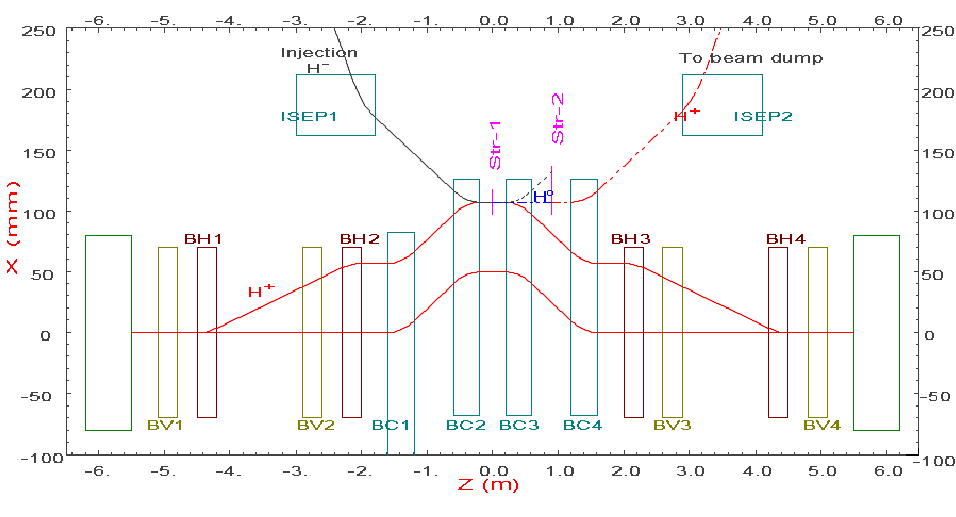}
\caption{\label{figinjlayout} The schematic layout of injection system}
\end{figure}

The RCS employs painting injection in the transverse direction to achieve uniform beam distribution and suppress the space-charge effects~\citep{ref3}~\citep{ref4}. Fig.~\ref{figinjlayout} shows the schematic view of the injection system. There are 12 bump magnets which are symmetrically distributed in one of the four straight sections in RCS. Eight bump magnets (BH1$\sim$4 and BV1$\sim$4) make a time dependent bump orbit during injection. The other four(BC1$\sim$4) bump magnets are used for generating horizontal chicane bump, which are excited by DC power supply, and the magnetic field is constant during the whole cycle. The edge focusing effects of the injection bump magnets cause beta-beating and break the fourfold symmetry of RCS, which will lead to tune shift and shrinkage of the acceptance. Meanwhile the strongest space charge arise during the first few milliseconds, which coincide with the strongest edge focusing effect of BC. The deterioration of the betatron motion stability due to edge focusing may cause extra beam loss when increasing the beam intensity. To achieve high intensity beam with low beam loss, it is important to compensate the edge focusing effects. Due to painting bump magnets only operating in the first 0.4 $ms$, only the BC is considered in this paper.  

In RCS, the main quadrupoles are excited by White resonant power supplies, and the exciting current cannot be arbitrarily programed. Generally, this kind of perturbation could be corrected by trim-quadrupole, which is powered by programmable power supply~\citep{ref5}~\citep{ref6}. However, there isn't trim-quadrupole available in CSNS/RCS. A new method based on harmonic injection was introduced to compensate the edge focusing effects of BC in CSNS/RCS. In CSNS/RCS, due to the eddy current effect and saturation of magnetic field, with the sinusoidal exciting current, the curve of magnetic field in a cycle is not sinusoidal. To compensate the magnetic field deviation, the harmonic injection was adopted, in which, the injected harmonic was obtained by a method based on the measured transfer function between exciting current and magnetic field~\citep{ref7}. By using this method, the curve of magnetic field in a cycle can be corrected very close to sinusoidal curve.

In this paper, the harmonic injection is adopted to compensate the beta-beating through modifying the curve of magnetic field of the main quadrupoles to the objective curve. 

The simulation study shows that the method could effectively compensate the beta-beating caused by the edge focusing of BC. The tune and the acceptance of the RCS could be adjusted to desired values by compensating.

\section{Beta-beating due to the edge focusing}

\begin{figure}[htb]
\includegraphics[width=8.5cm,height=5.5cm]{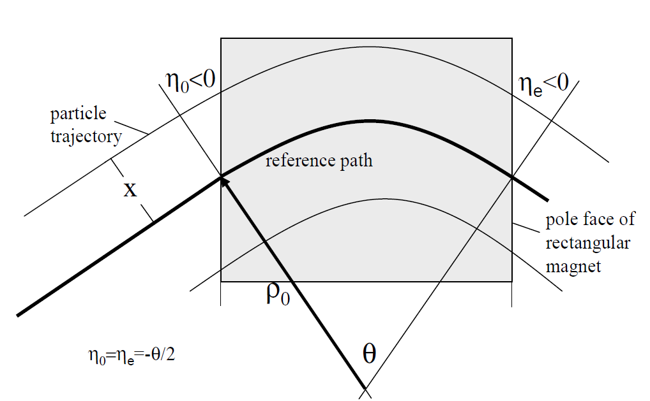}
\caption{\label{figedge} The diagrammatic sketch of the edge focusing}
\end{figure}

As the painting bump magnets only act on the beam during the injection procedure, here only 4 BC magnets are considered. Fig.~\ref{figedge} shows a diagrammatic sketch of the edge focusing effect. As shown in the figure, the edge angles at the entrance ($\eta_0$) and the exit ($\eta_e$) of the bump magnet were assumed to same. The transfer matrices of the horizontal bump magnet in the horizontal and vertical direction were shown in Eq~(\ref{eq1}) and Eq~(\ref{eq2}) respectively~\citep{ref8}. 

\begin{widetext}
\begin{eqnarray}
\label{eq1}
M_x = 
\begin{pmatrix}
1 & 0 \\ \dfrac{tan\frac{\theta}{2}}{\rho_0} & 1
\end{pmatrix}
\begin{pmatrix}
cos\theta & \rho_0sin\theta\\ 
-\dfrac{sin\theta}{\rho_0} & cos\theta
\end{pmatrix} 
\begin{pmatrix}
1 & 0 \\ \dfrac{tan\frac{\theta}{2}}{\rho_0} & 1
\end{pmatrix}
=
\begin{pmatrix}
1 & \rho_0sin\theta \\ 0 & 1
\end{pmatrix}
\\
\label{eq2}
M_y = 
\begin{pmatrix}
1 & 0 \\ -\dfrac{tan\frac{\theta}{2}}{\rho_0} & 1
\end{pmatrix}
\begin{pmatrix}
1 & \rho_0\theta\\ 
0 & 1
\end{pmatrix} 
\begin{pmatrix}
1 & 0 \\ -\dfrac{tan\frac{\theta}{2}}{\rho_0} & 1
\end{pmatrix}
=
\begin{pmatrix}
1-\theta tan\frac{\theta}{2} & \rho_0\theta \\ 
\dfrac{\theta tan^2\frac{\theta}{2}}{\rho_0}-\dfrac{2tan\frac{\theta}{2}}{\rho_0} & 1-\theta tan\frac{\theta}{2}
\end{pmatrix}
\end{eqnarray}
\end{widetext}

\noindent where $\theta$ is the bending angle of the bump magnet, $\rho_0$ is the bending radius, $M_x$ and $M_y$ are the transfer matrices of the horizontal bump magnet in horizontal and vertical direction respectively. And it is easily to prove that $\eta_0+\eta_e=-\theta$.

The edge focusing forces act as a quadrupole magnet, but as shown in Eq~(\ref{eq1}) the horizontal edge focusing effect is compensated by the intrinsic focusing property on the bending plane, while as shown in Eq~(\ref{eq2}) the vertical edge focusing effect remains. This typical property of the edge focusing effect is different from a normal quadrupole. Thus, the beta-beating caused by the edge focusing of the horizontal injection bump magnets occurs only on the vertical plane.

\begin{figure}[htb]
\includegraphics[width=7cm,height=7cm]{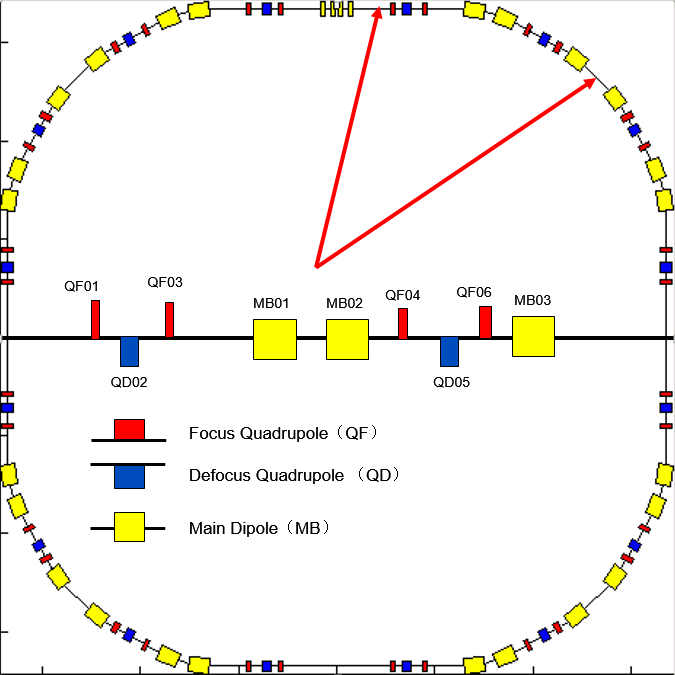}
\caption{\label{figlattice}(Color)The structure of the RCS.}
\end{figure}

As shown in Fig.~\ref{figlattice}, the RCS has a fourfold symmetric lattice, which is based on triplet cells, over its circumference of 227.92 m. In each super-period, the structure is mirror symmetry on the middle point, and there are 6 dipoles and 4 triplet cells. The total 48 quadrupoles are powered by 5 families power supply, which are resonant circuits. The resonant period of the power supply is 40 $ms$, and the beam is accelerated in the rising stage of the exciting current. 

\begin{figure}[htb]
\includegraphics[width=8cm]{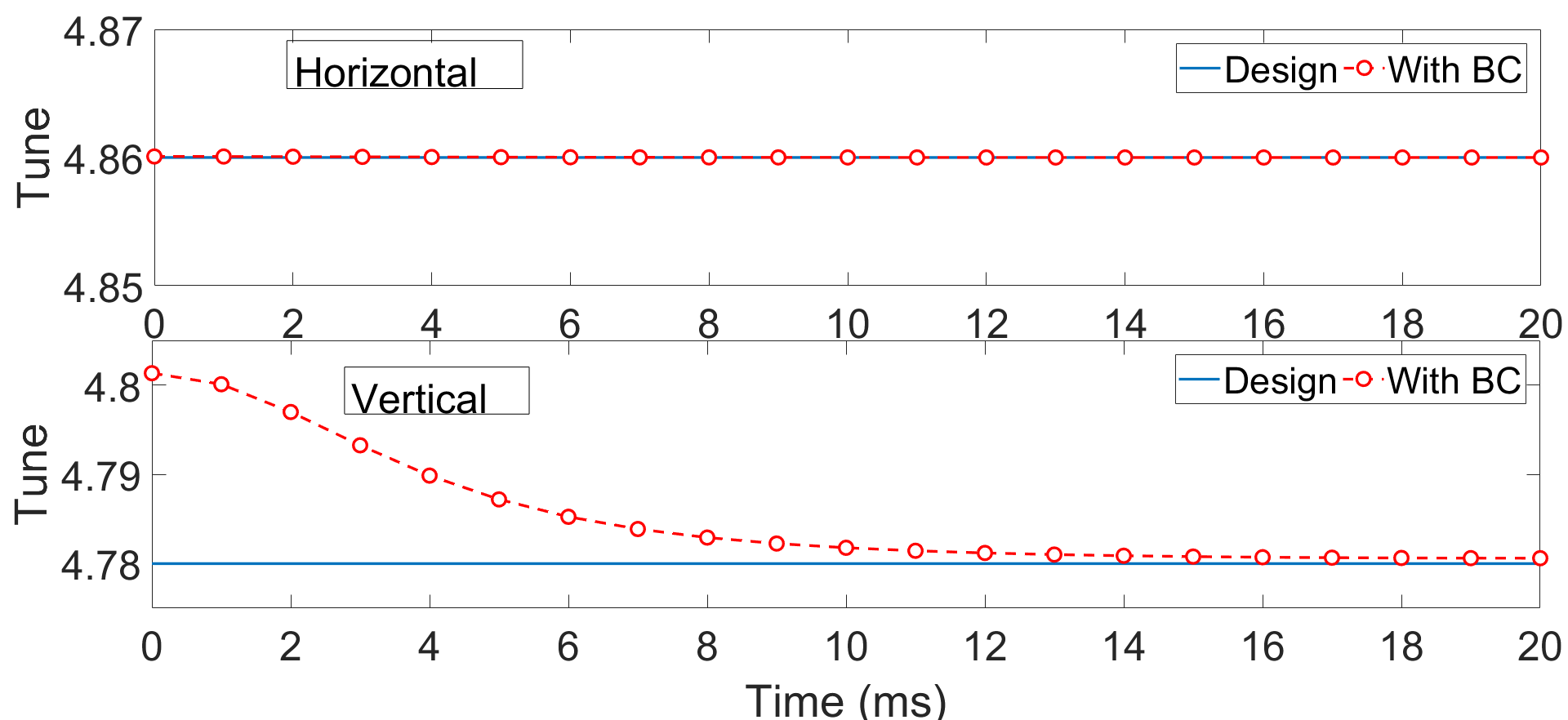}
\caption{\label{fig3}(Color)The tunes in one period with (red dash line with circles) and without (blue solid line) BC bump magnet.}
\end{figure}

\begin{figure}[htb]
\includegraphics[width=8cm,height=5cm]{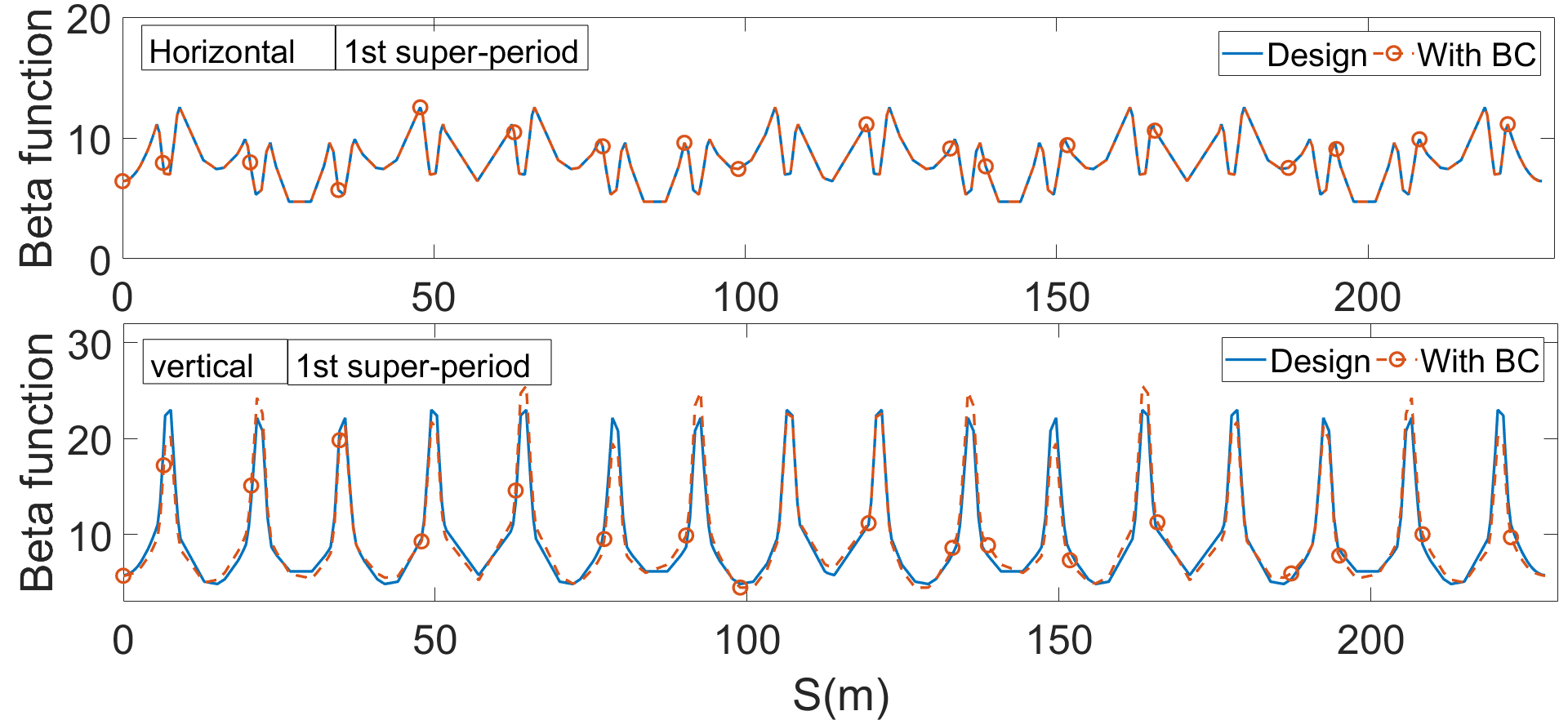}
\caption{\label{fig4}(Color)The beta functions in the first turn with (red dash line with circles) and without (blue solid line) BC bump magnet. }
\end{figure}

\begin{figure}[htb]
\includegraphics[width=8cm,height=5cm]{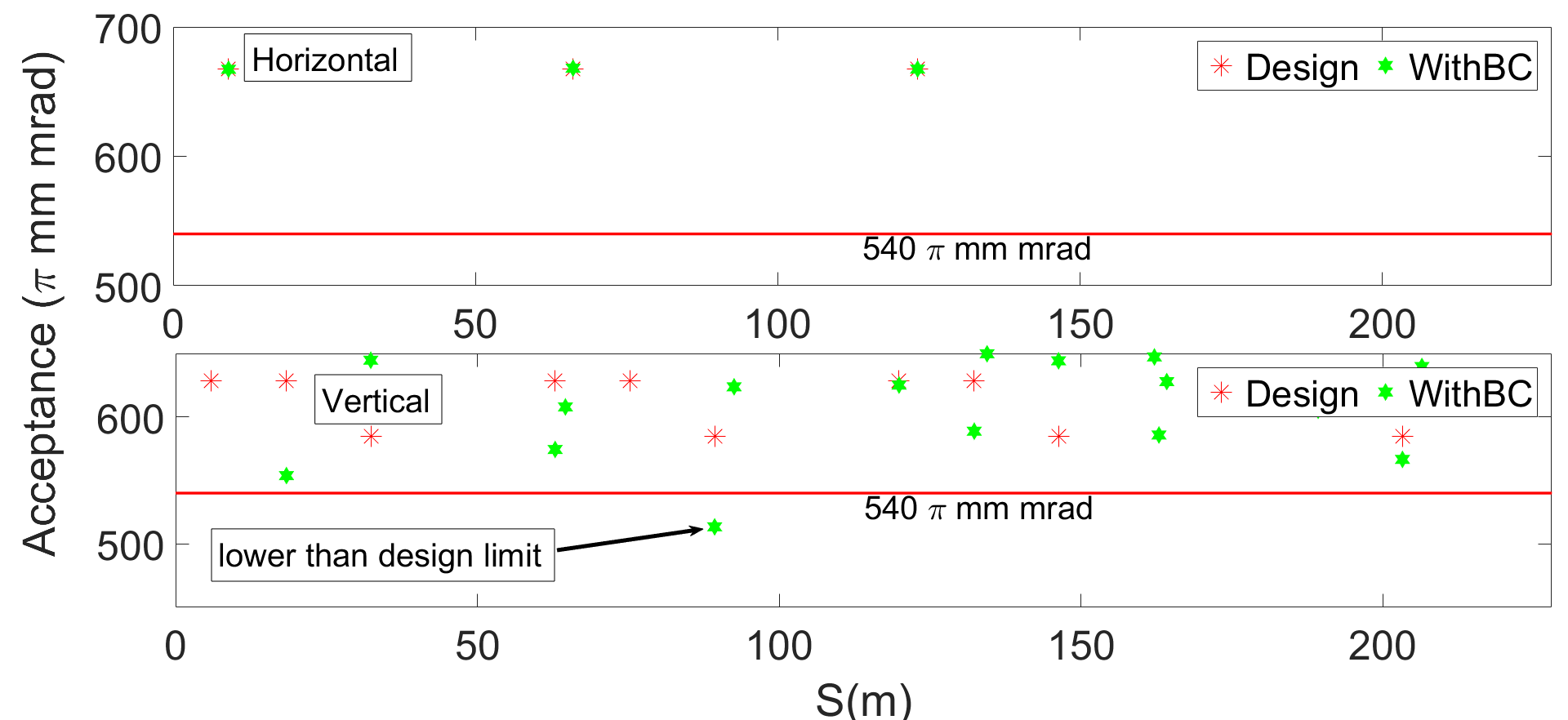}
\caption{\label{figacc}(Color)The acceptance of RCS with (green hexagram) and without (red asterisk) BC bump magnet. The red solid line is the design criteria. }
\end{figure}

MADX~\citep{ref9} was used to simulate the edge focusing effects on tune and beta function, while the acceptance could be calculated based on the beta function. The simulation results were shown in Fig.~\ref{fig3} and Fig.~\ref{fig4}, and the calculated acceptance were shown in Fig.~\ref{figacc}. The first figure shows the tunes with and without BC in one beam cycle, and the second figure shows the beta function in the first turn with and without BC respectively. In the vertical plane, the maximum shift of the tune is 0.02 and the maximum deviation of the beta function is 2.7 m, and one can see that the acceptance is lower than the design limit(540 $\pi mm mrad$) at some point.


\section{Compensation scheme for the edge focusing effect}

According to the tune shift calculation in section 2, as shown in Fig.~\ref{fig3}, the tune shift due to the edge focusing of BC is monotonically decreasing in the whole beam period of a RCS cycle, and the continuity of the tune shift makes it possible to compensate the edge focusing effect by the harmonic injection.
Different from the harmonic injection method used in the exciting curve correction for obtaining sinusoidal magnetic field in a cycle of RCS, the harmonic injection method used for compensating the edge focusing effects is to modify the curve of the magnetic field to an objective curve.

\begin{figure}[htb]
\includegraphics[width=8cm]{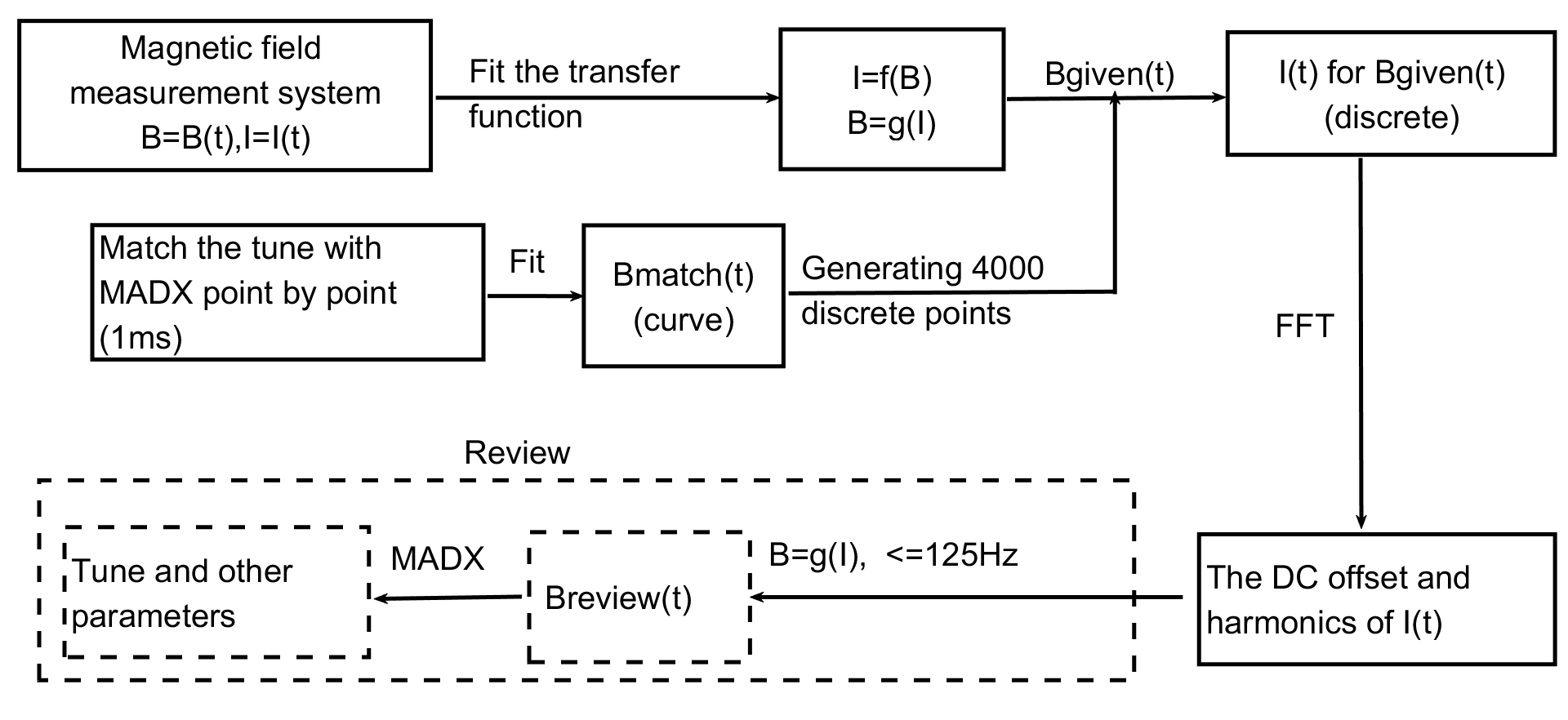}
\caption{\label{figcompenprocess} The flow process diagram of the compensation of edge focusing.}
\end{figure}

The detailed procedure, as shown in the Fig.~\ref{figcompenprocess}, can be divided into following steps:
\begin{enumerate}
\item calculating the beta-beating in a cycle of RCS. The calculation is performed every 1 ms, and a total of 21 points are chosen to perform the calculation in a cycle of 20 ms;
\item correcting the beta-beating point by point in a cycle, and fitting the curve of magnetic field for different families of quadrupoles;
\item calculating the injecting harmonic to be injected into exciting current;
\item re-calculating tunes based on the curve of magnetic field excited by the corrected curve of exciting current.
\end{enumerate}

The key step is to obtain the injecting harmonics. The new method based on the transfer function could easily get the corresponding injecting harmonics from the objective curve of the magnetic field. 

The transfer function is obtained through magnetic field measurement for AC mode. The magnetic field and exciting current at different time during the exciting current ramping are measured by using the harmonic coil measurement system~\citep{ref10}\citep{ref11}, then the transfer function $I=f(B)$ and $B=g(I)$ are fitted based on the measured data, where $I$, $B$ represent current and magnetic field respectively.

The objective curve of the magnetic field is obtained through matching the tune point by point and performing the fourier curve fitting on the matched results.

However, there are always many solutions of quadrupole settings for a specified tune, and the obtained curve of the quadrupole gradient may be unsmoothed if the matching is performed independently, and it is hard to derive the desired injecting harmonics with an unsmoothed gradient curve. So the matching process of each point should be interconnected. The relationship between tune shift and gradient deviation of quadrupole, as shown in Eq.~\ref{eqk}~\citep{ref12}, is helpful to get the smooth gradient curve of quadrupoles.
\begin{equation}
\label{eqk}
\Delta\nu=\frac{1}{4\pi}\sum\beta(s_1)k(s_1)\Delta s_1
\end{equation}
where $\beta$ is the beta function at the quadrupole, $k$ is the gradient error of the quadrupole, $s_1$ is the longitudinal position. Eq.~(\ref{eqk}) shows the betatron tune shift due to the gradient error, and this equation can be used to associate the matching process. According to the Eq.~(\ref{eqk}) the matching is performed at one time point only, and the magnetic fields at this time point could be obtained. Then the magnetic fields at other time points could be obtained by using the flowing derived equation,

\begin{equation}
\frac{\Delta\nu}{N}= \sum_{i=1}^{5}h_i\frac{k(s_i)}{N}
\end{equation}
where $h_i = \frac{1}{4\pi}\sum\beta(s_i)\Delta s_i$ is constant for the same quadrupole family, $N$ is the scale factor.

\begin{table*}[ht]
\caption{ \label{tabharm} The harmonics of the design current curve and the modified current curve.}

\begin{tabularx}{140mm}{|c|c|c|c|c|c|c|c|c|c|c|c|}
\hline\hline
\multicolumn{2}{|c|}{\multirow{2}{*}{\diagbox{Harmonic}{Magnet}}} & \multicolumn{2}{|c|}{Q206A} & \multicolumn{2}{|c|}{Q272} &
 \multicolumn{2}{|c|}{Q206B} & \multicolumn{2}{|c|}{Q222} &\multicolumn{2}{|c|}{Q253}\\
\cline{3-12}
 \multicolumn{2}{|c|}{~}& design & modify& design & modify& design & modify& design & modify& design & modify\\
\hline
\multicolumn{2}{|c|}{DC($A$)}  & 720.7 & 715.9 &794.1 & 793.6 &642.9  & 647.5 &618.2 & 621.9 &747.6  &743.4\\
\hline
\multirow{2}{*}{25Hz}&Amp($A$) & 531.4 & 534.6 &568.5 & 569   &466.3  & 461.6 &446.4 & 441   &538.7  &545.2\\
\cline{2-12}
               & Phase($\pi$)  & 1.5   & 1.5   &1.5   & 1.5   & 1.5   & 1.5   & 1.5  & 1.5   & 1.5   & 1.5\\
 \hline
\multirow{2}{*}{50Hz}&Amp($A$) & 15.24 & 13.11 &8.57  & 8.298 &8.211  & 10.19 &4.537 & 6.978 &8.876  &6.196\\
\cline{2-12}
               & Phase($\pi$)  &0.145  & 1.59  &0.79  & 0.78  &0.58   & 0.62  &0.52  & 0.7   &0.67   &0.49\\
 \hline
\multirow{2}{*}{75Hz}&Amp($A$) &3.727  & 4.273 &7.745 &7.576  &4.813  & 6.015 &1.767 & 4.529 &6.738  &2.957\\
\cline{2-12}
               & Phase($\pi$)  &0.7    & 1.4   &1.53  & 1.53  &1.445  & 1.466 &1.308 & 1.439 &1.43   &1.34\\
 \hline
\multirow{2}{*}{100Hz}&Amp($A$)&2.623  & 2.788 &3.744 &3.583  &1.182  & 2.394 &0.2227& 2.381 &2.454  &0.4168\\
\cline{2-12}
                & Phase($\pi$) &0.85   & 0.81  &0.02  &0.02   &0.23   & 0.15  &1.74  & 0.01  &  0    &1.05\\
 \hline
\multirow{2}{*}{125Hz}&Amp($A$)&2.901  & 2.05  &1.495 &1.407  &0.3251 & 0.7437&0.29  & 1.109 &0.9701 &0.6434\\
\cline{2-12}
                 & Phase($\pi$)&1.7    & 1.63  &0.46  &0.46   &1.99   & 0.34  &0.054 & 0.41  &0.294  &1.711\\
\hline\hline
 
\end{tabularx}
\end{table*}

\begin{figure}[h]
\includegraphics[width=8cm]{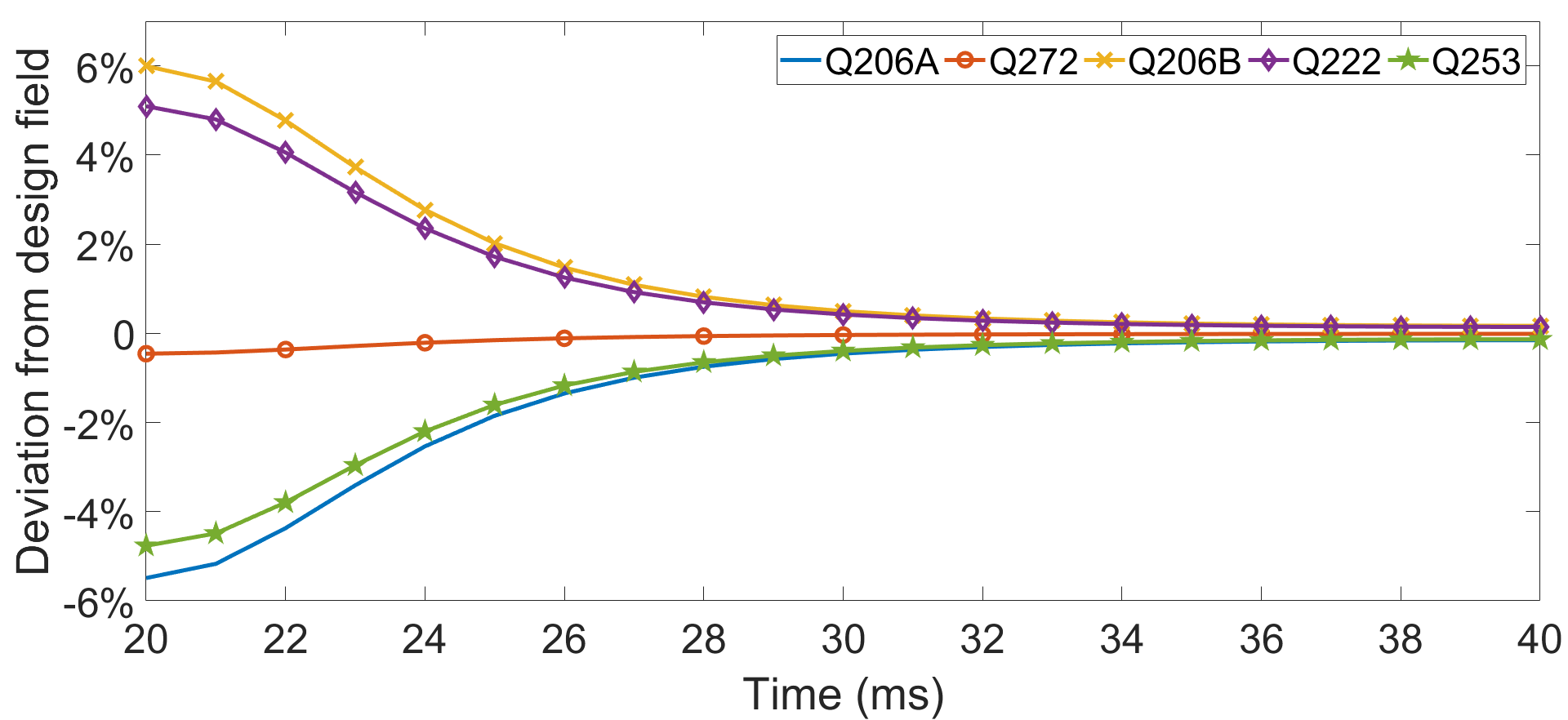}
\caption{\label{figfieldchange}(Color)The modified magnetic field curves, which adjust the tune to the design values, relative to the design magnetic field. Five quadrupole families are Q206A(blue solid line), Q272(red solid line with circles), Q206B(yellow solid line with cross), Q222(purple solid line with diamond) and Q253(green solid line with pentagram) respectively}
\end{figure}

\begin{figure}[h]
\includegraphics[width=8cm]{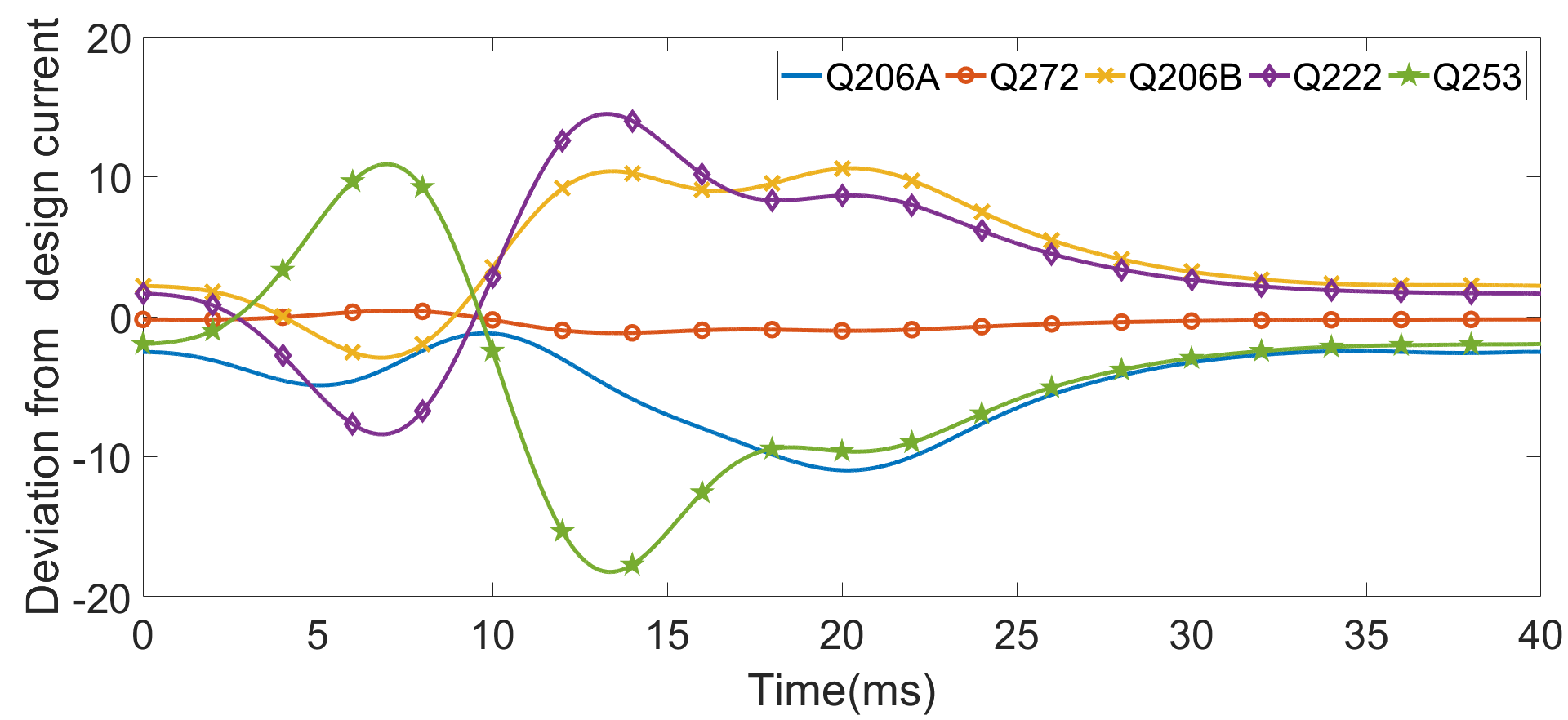}
\caption{\label{figcurrchange}(Color)The exciting current curve, which are corresponding to Fig.~\ref{figfieldchange}, relative to the design current curve. The markers are same as Fig.~\ref{figfieldchange}}
\end{figure}

The MADX and MATLAB~\citep{ref13} are used to perform the calculation and simulation. The matched results are a series of discrete points for each quadrupole family, and the objective curve of the magnetic field $B_{match}(t)$ are obtained by using fourier curve fitting. As shown in Fig.~\ref{figfieldchange}, one can see that the obtained magnetic field curve relative to the design magnetic field is smooth as expected. And as shown in Fig.~\ref{figcurrchange}, the corresponding exciting current $I(t)$ are derived from  $B_{given}$, which are 4000 discrete points generated from $B_{match}$, and the transfer function $I=f(B)$. The DC offset and harmonics of the derived exciting current, as shown in Table.~\ref{tabharm}, are obtained through FFT to $I(t)$. In practice, the magnetic field curve can be accurately control to the objective pattern $B_{match}(t)$ by injecting the obtained DC offset and time harmonics of the current into the resonant circuits. 

In the simulation study, in order to verify the compensation scheme, the curve of the exciting current which are generated from DC offset and time harmonics could be reconverted to magnetic field curve $B_{review}(t)$ through the transfer function $B=g(I)$. Then the tunes and other parameters could be obtained by applying $B_{review}(t)$ to the MADX model.

\begin{figure}[htb]
\includegraphics[width=8cm,height=5cm]{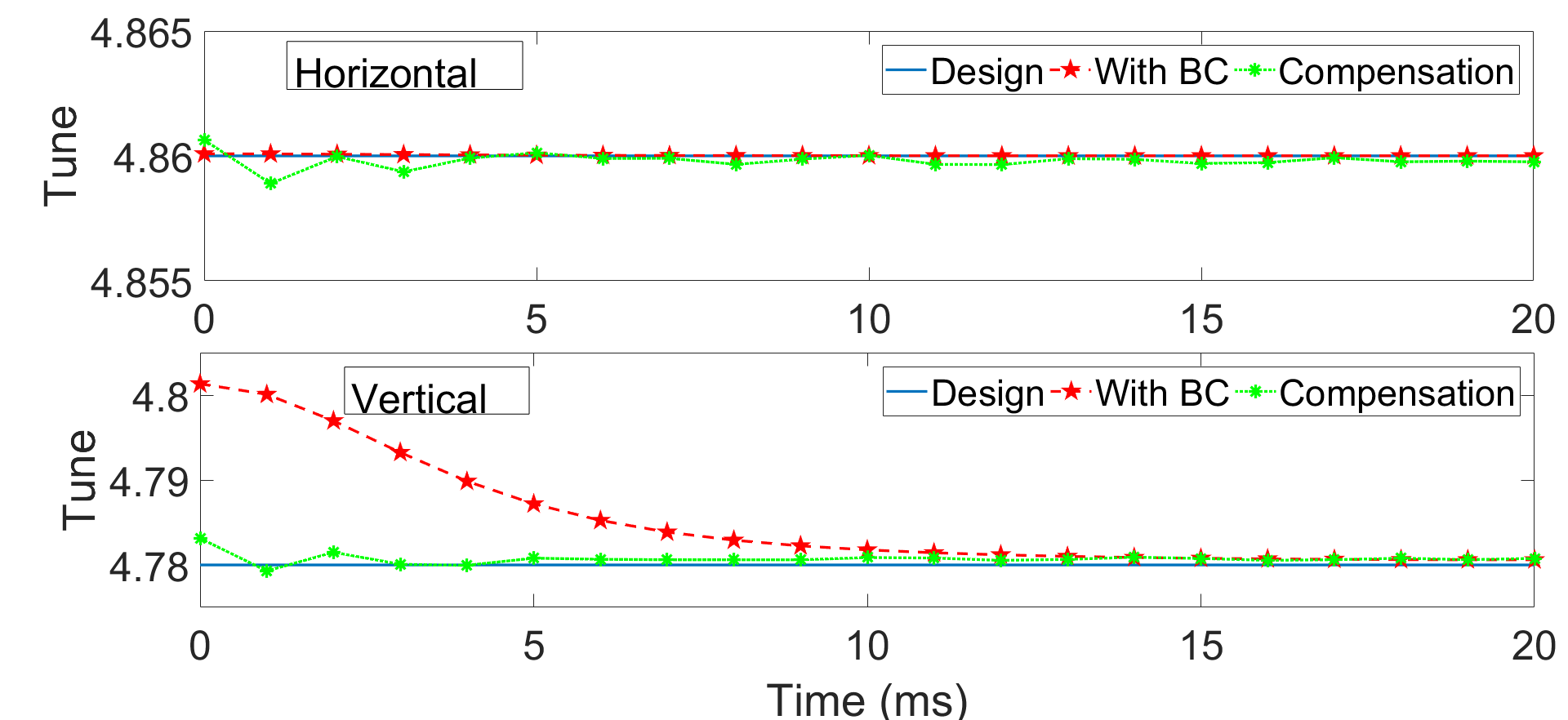}
\caption{\label{figcompentune}The tunes in one period with (red dash line with pentagram) and without (blue solid line) BC bump magnet compared with the tune after compensation(green dot line with asterisk).}
\end{figure}

\begin{figure}[htb]
\includegraphics[width=8cm,height=5cm]{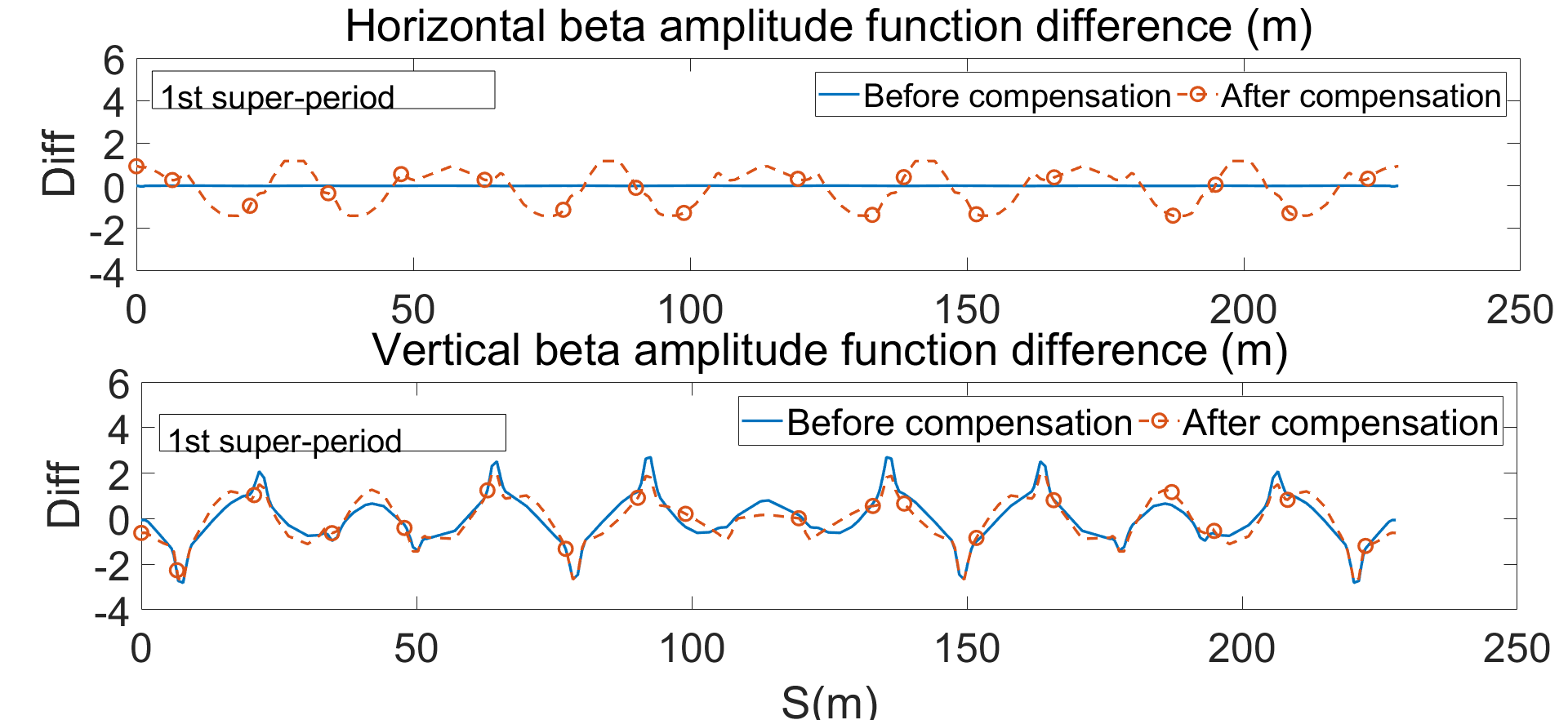}
\caption{\label{figcompenbeta} The beta function against the design value in the first turn with BC bump magnet before (blue solid line) and after (red dash line with circles) compensation respectively.}
\end{figure}

\begin{figure}[htb]
\includegraphics[width=8cm]{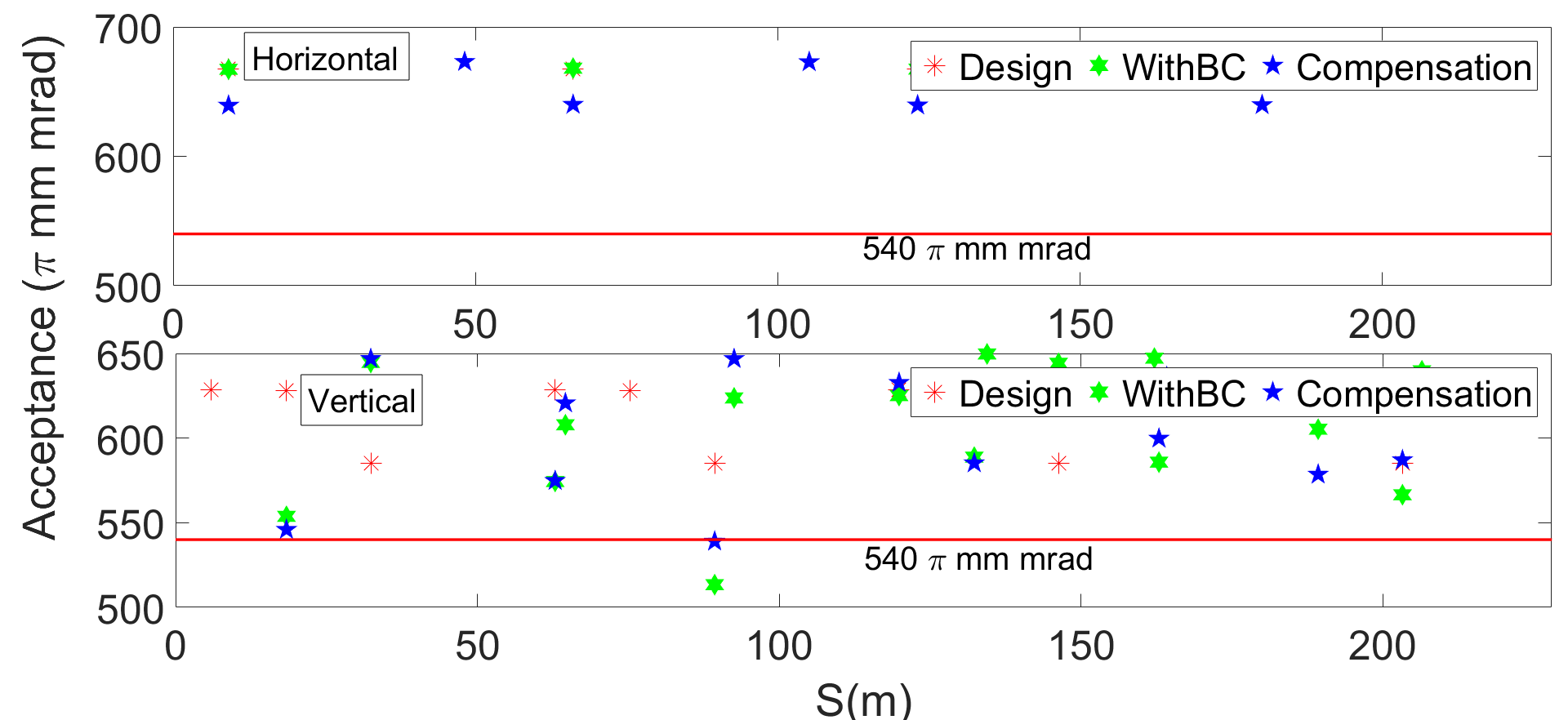}
\caption{\label{figacceptance} The acceptance of the RCS after compensation compared with the acceptance with and without (Design) BC bump magnet.}
\end{figure}

As shown in Fig.~\ref{figcompentune}, the tunes in a cycle after compensation are very close to the design value, in which the maximum deviation is decreased from 0.02 to 0.002. Fig.~\ref{figcompenbeta} shows the beta function relative to the design value in the first turn before and after compensation. The maximum deviation relative to the design value in the vertical plane is decreased from 2.7 m to 1.98 m, and as shown in the Fig.~\ref{figacceptance} the acceptance of RCS after compensation was controlled above the design requirement of 540 mmmrad.

\vspace{3mm}
\section{Conclusion}

The edge focusing effects of the bump magnets give rise to tune shift in the whole beam period of a RCS cycle, and the shrinkage of the acceptance may result in extra beam loss when increasing beam intensity. For the chicane bump magnet BC, a new compensation method was introduced in this paper, which is based on the harmonic injection to the main quadrupoles. The simulation study show the validity and effectiveness of the new method. The tune in the whole RCS cycle could be adjusted to the designed value and the acceptance of RCS also could be recovered to meet the design specification.

In compensating the edge focusing effect induced tune shift by using the method of harmonic injection, the key technology is to calculate the injected harmonics. With the aid of comprehensive measured data of magnetic field in AC mode, by using the transfer function method, it is easy to get the injected harmonics. In the commissioning and operation of a RCS, always some tune deviation from the design value during a RCS cycle, and this method can also be used to correct this kind of tune or beta function deviation.

\section{Acknowledgements}
The authors want to thank M.Y.Huang, Y.W.An and other CSNS colleagues for the discussion and consultations.

\clearpage
\end{CJK}
\end{document}